\documentclass[prb,floatfix,twocolumn,showpacs,superscriptaddress]{revtex4}
\usepackage{dcolumn}% Align table columns on decimal point
\usepackage{epsfig}
\begin{document}
\newcommand{\bc}{\begin{center}}
\newcommand{\ec}{\end{center}}
\newcommand{\be}{\begin{equation}}
\newcommand{\ee}{\end{equation}}
\newcommand{\beqn}{\begin{eqnarray}}
\newcommand{\eeqn}{\end{eqnarray}}

\title{Magnetic ordering in a doped frustrated spin-Peierls system}

\author{Nicolas Laflorencie}
\affiliation{Laboratoire de Physique Th\'eorique, CNRS-UMR5152
Universit\'e Paul Sabatier, F-31062 Toulouse, France}

\author{Didier Poilblanc}
\affiliation{Laboratoire de Physique Th\'eorique (CNRS-UMR5152),
Universit\'e Paul Sabatier, F-31062 Toulouse, France}

\author{Anders W.~Sandvik}
\affiliation{Department of Physics, {\AA}bo Akademi University,
Porthansgatan 3, FIN-20500 Turku, Finland}

%\homepage{http://ww3-phystheo.ups-tlse.fr/~didier}

\date{\today}

\begin{abstract}

Based on a model of a quasi-one dimensional 
spin-Peierls system doped with non-magnetic
impurities, an effective two-dimensional
Hamiltonian of randomly distributed S=1/2 spins 
interacting via long-range pair-wise interaction is
studied using a stochastic series expansion 
quantum Monte Carlo method. The susceptibility shows Curie-like behavior 
at the lowest temperatures reached although the staggered magnetisation
is found to be finite for $T\rightarrow 0$. The doping dependance of 
the corresponding three-dimensional N\'eel temperature is also
computed.

\end{abstract}

\maketitle

%============ BODY OF PAPER ===================================

Quasi one-dimensional (1D) quantum antiferromagnets exhibit fascinating 
magnetic properties at low temperatures. Some inorganic compounds, such as 
the germanate oxide CuGeO$_3$~\cite{Hase93} and the vanadate oxide 
LiV$_2$O$_5$~\cite{LiV2O5} are excellent realizations of
weakly interacting frustrated spin-1/2 chains. A spin-Peierls (SP)
transition to a gapped dimerised ground state (GS) has been
seen experimentally in CuGeO$_3$~\cite{Hase93}, and theoretical
calculations~\cite{Becca2002} point toward a similar scenario in
LiV$_2$O$_5$. Doping with non-magnetic dopants is realized
experimentally in CuGeO$_3$ by substituting a small fraction of
copper atoms by zinc (or magnesium) atoms~\cite{Zn_CuGeO}. An
intriguing low-temperature phase where antiferromagnetism
coexists with the SP dimerisation was observed~\cite{Zn_CuGeO}.
Similar doping-induced antiferromagnetic (AF) ordering has also been
observed in the interacting dimer compound TlCuCl$_3$~\cite{doped_dimers}.
Theoretically, each dopant is expected to release a soliton which
can be viewed as a single unpaired spin separating two dimer
configurations~\cite{Sorensen98}, hence leading to a rapid
suppression of the spin gap under doping~\cite{Martins96}. In an
idealized {\it spontaneously} dimerised spin chain the soliton
would not be bound to the dopant~\cite{Sorensen98}. The physical
picture is in fact completely different: A static {\it bulk} dimerisation 
is enforced by, e.g., couplings to the three-dimensional (3D) lattice, 
thus generating an attractive potential between the soliton and the dopant
\cite{Sorensen98,Nakamura99}.

The dopant-soliton confinement mechanism is responsible for the
formation of local S=1/2 magnetic
moments~\cite{Sorensen98,Normand2002}. Within a realistic model
including an elastic coupling to a two-dimensional (2D)
lattice~\cite{Hansen99}, it was shown that these effective spins
experience a non-frustrated interaction that could lead at $T=0$
to a finite staggered magnetization~\cite{Dobry98}. Recently,
similar conclusions were reached using a model with purely
magnetic interactions including a four-spin exchange
coupling~\cite{ourprl03}. In this Letter, we analyze the formation
of the dopant-induced AF order which coexists
with the SP dimerisation.  After using exact diagonalisation (ED)
of small clusters (following Refs.~\cite{Dobry98,ourprl03}) to
construct an effective diluted S=1/2 model, we take advantage of the
non-frustrated character of the resulting Hamiltonian to perform extensive
state-of-the-art stochastic series expansion (SSE) quantum Monte
Carlo (QMC) simulations on 2D lattices as large as $288 \times 288$
with up to $N_s=576$ (dopant) spins and down to temperature as low
as $T=1/\beta=2^{-14}$ (or $2^{-18}$ for $N_s=256$). 
The uniform susceptibility 
is shown to exhibit a Curie-like behavior although 
the $T=0$ and infinite size extrapolated staggered magnetization 
is found to be finite down to the smallest dopant concentrations $x$ available.
The N\'eel temperature (assuming a small 3D coupling)
is also computed versus $x$ and compared to experiments. 

\begin{figure}
\bc \epsfig{file=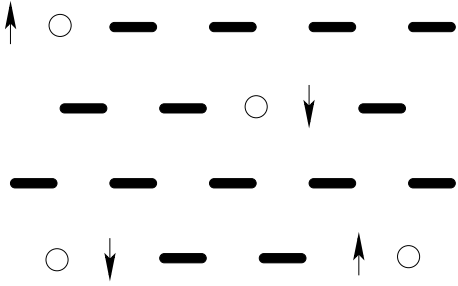,width=6cm} 
\caption{Schematic picture
of a doped SP system. Thick bonds correspond to dimers, and 
non-magnetic dopants (released spin-$\frac{1}{2}$)
are represented by open circles (arrows).}
\label{fig:Pict} \ec
\end{figure}

We start with the microscopic Hamiltonian of a 2D array of 
coupled frustrated spin-$\frac{1}{2}$ chains and we 
summarize the procedure followed in refs.~\cite{Dobry98,ourprl03}
to derive an effective Hamiltonian.
\begin{eqnarray}
\label{hamilQ1D.MF} H =\sum_{i,a}[J(1&+&\delta_{i,a})\, {\bf
S}_{i,a}\cdot{\bf S}_{i+1,a}~\nonumber \\
&+& \alpha J\, {\bf S}_{i,a}\cdot {\bf S}_{i+2,a}
+h_{i,a}S_{i,a}^z]\, ,
\end{eqnarray}
where the $i$ and $a$ indices label the $L$ sites and $M$ chains
respectively. The energy scale is set by the exchange coupling
along the chain ($J=1$) and $\alpha$ is the relative magnitude of
the next nearest neighbor frustrating magnetic coupling.
Dopants are simply described as randomly located inert sites $(i,a)$
(see Fig.~\ref{fig:Pict}) where ${\bf S}_{i,a}={\bf 0}$ is set in 
Eq.~(\ref{hamilQ1D.MF}). Small
inter-chain couplings are included in a mean-field
self-consistent treatment assuming,
\begin{eqnarray}
\label{self} h_{i,a}&=&J_{\bot}(\langle S_{i,a+1}^z
\rangle+\langle S_{i,a-1}^z \rangle)\, ,\\
\delta_{i,a}\!&=&\!\frac{J_4}{J}\lbrace\langle {\bf
S}_{i,a+1}\cdot{\bf S}_{i+1,a+1} \rangle\!+\!\langle {\bf
S}_{i,a-1}\cdot{\bf S}_{i+1,a-1} \rangle\rbrace .\label{self2}
%\nonumber.
\end{eqnarray}
While the first term accounts for first-order effects in the
inter-chain magnetic coupling $J_\perp$, the second term might
have multiple origins; although a four-spin cyclic exchange
mechanism provides the most straightforward derivation of
it~\cite{ourprl03}, at a qualitative level, $J_4$ can also mimic
higher-order effects in $J_\perp$~\cite{Sushkov} or the coupling
to a 2D (or 3D) lattice. In that case, due to a magneto-elastic
coupling, the modulations $\delta_{i,a}$ result 
from small displacements of the ions. The elastic 
energy is the sum of a local term
$\frac{1}{2}K_\parallel\sum_{i,a}\delta_{i,a}^2$ and an
inter-chain contribution $K_\perp\sum_{i,a}
\delta_{i,a}\delta_{i,a+1}$ of electrostatic
origin~\cite{Affleck}. In that case Eq.~(\ref{self2}) is replaced
by~\cite{Hansen99},
$
K_\parallel\delta_{i,a}+K_\perp(\delta_{i,a+1}+\delta_{i,a-1})=
J\langle {\bf S}_{i,a}\cdot{\bf S}_{i+1,a} \rangle
$, 
giving very similar
results~\cite{note1} so that we shall restrict to
Eq.~(\ref{self2}) here.

%%%%%%%%%%%%%%%FIG:PICT%%%%%%%%%%%%%%%%%%%%%%%%%%%
%%%%%%%%%%%%%%%%%%%%%%%%%%%%%%%%%%%%%%%%%%%%%%%%%%

By breaking a dimer,
each dopant  releases a soliton carrying a spin
1/2. As shown previously~\cite{ourprl03}, $J_4$ (or equivalently
$K_\perp$ in the alternative model~\cite{Hansen99}) leads to a
confinement of the moment to the dopant with a localization length
$\sim J_{4}^{-\eta}$. At temperatures lower than
the spin gap, these effective spins dominate the physics (see
Fig.\ref{fig:Pict}) and a low-temperature description of the doped
system can be obtained using an effective model including only
(long-range) pair-wise interaction between these local moments.
The coupling $J^{\rm eff}$ between two effective
spins at arbitrary relative distance is
computed by ED as the energy difference between their singlet and
triplet configurations~\cite{Dobry98,ourprl03}. The sign 
of this interaction (i.e. its ferromagnetic or AF
nature) depends on whether the two
dopants lie on the same or opposite sublattices~\cite{note2}, 
so that an overall AF ordering is favored.

In order to
derive an analytic expression for $J^{\rm eff}$ valid
at long distances, we fit the numerical ED data (restricted to short and 
intermediate length scales).
A long-range {\it non-frustrated} Heisenberg 
model of diluted effective spin-$\frac{1}{2}$ can then be defined,
\be
{\mathcal{H^{\rm{eff}}}}
=\sum_{{\bf r}_1,{\bf r}_2}
\epsilon_{{\bf r}_1} \epsilon_{{\bf r}_2}J^{\rm eff}({\bf r}_1-{\bf r}_2)
{\bf S}_{{\bf r}_1}\cdot {\bf S}_{{\bf r}_2} ,
\label{EffHam} 
\ee 
with $\epsilon_{\bf r}=1$ ($0$) with probability $x$ ($1-x$),
where $x$ is the dopant concentration. Such a model can be 
studied by QMC on $L_x \times L_y$ clusters much larger than 
those accessible to ED~\cite{Dobry98} and {\it at all temperatures}.
Using five 
phenomenological parameters, two energy scales and three
length scales, simple expressions fit the ED data for
a wide range of (physical) parameters. Naturally, one has to 
distinguish four
cases depending on whether the dopants are located on the same
($\Delta a=0$) or on different chains, and whether they are
located on the same or on different sublattices. When $\Delta a=0$
(same chain), $J^{\rm eff}$ approximately fulfills $J^{\rm eff}
(\Delta i,0)=J_0(1-\Delta i/\xi_{\parallel}^0)$ for $\Delta i$
even and $\Delta i<\xi_{\parallel}^0$ and otherwise $J^{\rm eff}(\Delta
i,0)=0$. For dopants located on different chains and on the same 
sublattice ($\Delta i+\Delta a$ even) one has,
$
J^{\rm eff}(\Delta i,\Delta a)=-J_{0}^\prime\exp(-\frac{\Delta
i}{\xi_{\parallel}}) \exp(-\frac{\Delta a}{\xi_{\perp}})\, ,
$
%\ee
while, if the dopants are on opposite sublattices,
$
J^{\rm eff}(\Delta i,\Delta a)=
J_{0}^\prime\frac{\Delta i}{2\xi_{\parallel}}\exp(-\frac{\Delta a}
{\xi_{\perp}}) 
$
for $\Delta i \le 2\xi_{\parallel}$ and 
$
J^{\rm eff}(\Delta i,\Delta a)=J_{0}^\prime \exp(-\frac{\Delta
i-2\xi_{\parallel}}{\xi_{\parallel}})\exp(-\frac{\Delta a}{\xi_{\perp}}), 
$
for $\Delta i > 2 \xi_{\parallel}$. The fitting parameters used here
for $\alpha=0.5$, $J_\perp=0.1$ and $J_4=0.08$ are $J_0=0.52$, 
$J_{0}^\prime=0.3$,
$\xi_{\parallel}^0=17.3$, $\xi_{\parallel}=2.5$ and $\xi_{\perp}=1$. 
We stress that the alternating sign of $J^{\rm eff}$
is a crucial feature of the interaction
which guarantees the absence of frustration.
Note also that the magnitude of $J^{\rm eff}$
shows a unique exponential behavior 
$
\propto\exp(-\frac{\Delta i}{\xi_{\parallel}} 
     -\frac{\Delta a}{\xi_{\perp}})
$ 
at long distance although its short distance behavior
is more complicated (but probably not relevant). 
The distribution of these coupling contains a very large density of couplings of small magnitudes. 

We study the effective Heisenberg model using the SSE method
\cite{Sandvik98} to investigate GS as well as
finite $T$ properties. In this approach, the interactions are sampled
stochastically, and for a long-ranged interaction the computational
effort is then reduced from $\sim N_s^2$ to $N_s\ln{(N_s)}$
\cite{Sandvik03}. In order to accelerate the convergence of the
simulations at the very low temperatures needed to study the ground
state, we use a $\beta$-doubling scheme \cite{SandPerc} where the
inverse temperature is successively increased by a factor $2$. 
Comparing results at several $\beta =2^n$, one can subsequently 
check that the $T\to 0$ limit has been reached.

%%%%%%%%%%%%%%%%FIG:GS.CVG.256%%%%%%%%%%%%%%%%%
\begin{figure}
\bc \epsfig{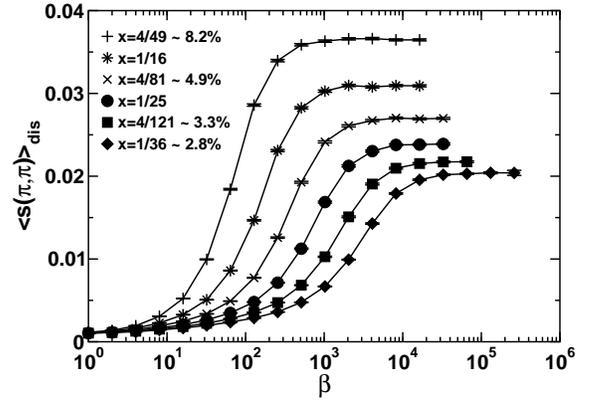} \caption{Staggered
magnetic structure factor per site vs
inverse-temperature $\beta$ computed using a $\beta$-doubling
scheme and averaged over $1000$ to $2000$ samples. Results for
$N_s =256$ spins on $56\times 56$, $64 \times 64$, $72\times 72$,
$80\times 80$, $88\times 88$ and $96\times 96$ (from top to
bottom) lattices correspond to the concentrations $x$ indicated on
the plot.} \label{fig:GS.CVG.256} \ec
\end{figure}
%%%%%%%%%%%%%%%%%%%%%%%%%%%%%%%%%%%%%%%%%%%%

The AF ordering instability is signalled by the divergence with
system size of the staggered structure factor, 
\be \label{StgStc}
S(\pi,\pi)=\frac{1}{L_xL_y} \langle (\sum_{i} (-1)^{i}
S_{i}^{z})^2\rangle \, .
\ee 
Note that within our effective model approach, only the sites carrying a 
"dopant spin" contribute to this sum. It is convenient to normalized $S$ 
with respect to the number of sites, i.e. to define a staggered structure
factor per site; $s(\pi,\pi)=S(\pi,\pi)/L_x L_y$. In an ordered AF state,
$s(\pi,\pi)$ should converge, with increasing size, to a non-zero 
value $< 1/4$. The (finite size) sublattice magnetization $m_{\rm AF}$ can 
then be obtained by averaging $s(\pi,\pi)$ over a 
large number of dopant distributions, i.e. 
$(m_{AF})^{2} = 3\langle s(\pi,\pi)\rangle_{\rm {dis}}$, where
the factor 3 comes from the spin-rotational invariance~\cite{Reger88} 
and $\langle\ldots\rangle_{\rm {dis}}$ stands for the disorder average. 
The staggered magnetization per dopant is then simply
$m_{\rm {spin}}=m_{\rm AF}/x$. We have checked that extrapolations to 
the thermodynamic limit using different aspect ratios $L_y/L_x$ give similar 
results and, here,  we only report data for $L_y=L_x=L$.

%%%%%%%%%%%FIG:FSC%%%%%%%%%%%%%%%%%
\begin{figure}
\bc \epsfig{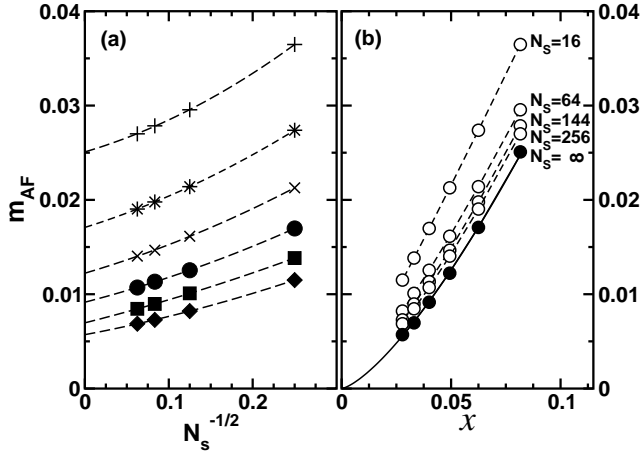} \caption{Staggered
magnetization per site. Disorder average has been done
over at least $2000$ samples. (a) Finite size extrapolations 
(see text) at fixed doping $x$. The symbols are identical to 
Fig.\protect\ref{fig:GS.CVG.256} for $x$ 
varying from $\sim 2.8\%$ to $\sim 8.2\%$. (b) Doping dependance
of $m_{\rm AF}$ for various numbers of spins and in the thermodynamic 
limit (full symbols). 
} \label{fig:mAF} \ec
\end{figure}
%%%%%%%%%%%%%%%%%%%%%%%%%%%%%%%%%%%

Since, strictly speaking, in 2D the divergence of $S(\pi,\pi)$
occurs only at $T=0$ ($\beta=\infty$) it is appropriate to first
extrapolate the finite size numerical data to $T=0$. As shown in
Fig.~\ref{fig:GS.CVG.256} the staggered structure factor saturates
at sufficiently low temperature and the GS value of $m_{AF}$
(averaged over disorder) can be safely obtained. Then, using a
polynomial fit in $1/\sqrt{N_{s}}$ (order 2 is sufficient) an
accurate extrapolation to the thermodynamic limit, $N_{s} \to
\infty$ (or $L\to\infty$ at constant $x$), is performed as shown
in Fig.\ref{fig:mAF}(a). The doping dependance of the extrapolated
$m_{AF}$ is given in Fig.\ref{fig:mAF}(b). Note that our 
results, although consistent with previous ($T=0$) extrapolations attempted
on small clusters~\cite{Dobry98}, are far more accurate due to the
use of much larger systems. We have tested various fits to the data. 
Assuming a power law $\propto x^{\mu}$, the best fit (solid line in Fig.\ref{fig:mAF}(b)) gives an 
exponent $\mu\simeq 1.38 >1$.
However, the alternative behavior
$a_1 x+ a_2 x^2$ would only be distinguishable at even 
smaller $x$.

%%%%%%%%%%%%%%%FIG:CurieT%%%%%%%%%%%%%%%%%%%%%%%%%%%
\begin{figure}
\bc 
\epsfig{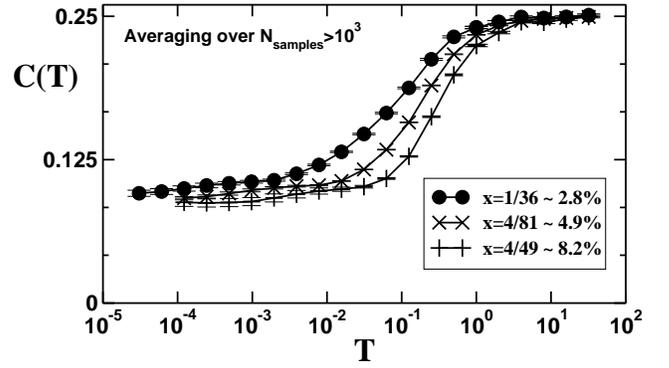} 
\caption{Curie constant $\chi \times T$ vs $T$ shown for 
$N_s =144$ spins and $3$ different concentrations $x$
(as shown on plot).}
\label{fig:Curie}
\ec
\end{figure}
%%%%%%%%%%%%%%%%%%%%%%%%%%%%%%%%%%%%%%%%%%%%%%%%%%

We have computed the uniform 
susceptibility $\chi(T)$ for a wide range of temperatures.
Results for the Curie constant $C(T)=T\chi(T)$ 
are shown in Fig.~\ref{fig:Curie}. At the highest temperatures
the effective dopant spins behave as free spins while at low temperature
we observe a new Curie-like behavior with a reduced Curie constant
$\sim 1/12$. Although here the 2D system orders 
at $T=0$ (as proven above) 
this behavior agrees with a qualitative argument 
by Sigrist and Furusaki~\cite{note2} based on the formation of large 
spin clusters. A detailled analysis of a
low-temperature scaling regime similar to the one
observed in random ferromagnetic-antiferromagnetic 
spin chains~\cite{F-AF} will be reported 
elsewhere~\cite{scaling}.  

\begin{figure}
\bc \epsfig{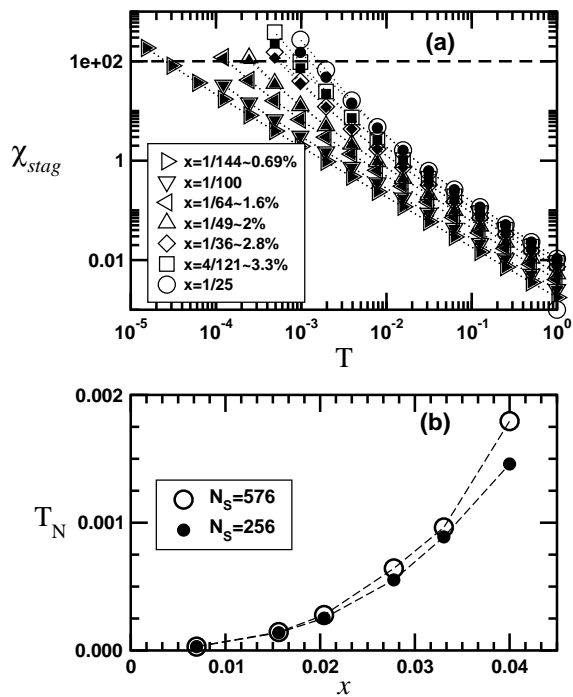} \caption{(a)
Staggered susceptibility of a 2D layer vs temperature (using
log-log scales) for $N_s=256$ (full symbols) and $N_s=576$ (open
symbols) spins. Concentrations $x$ are shown on the
plot (similar symbols as in Figs.~\protect\ref{fig:GS.CVG.256}
and \protect\ref{fig:mAF}). $\lambda_{3D}^{-1}=100$ is shown by the dashed line. 
(b)
N\'eel temperature vs dopant concentration $x$ for a 3D RPA
inter-plane coupling $\lambda_{3D}=0.01$ and for $N_s=256$ and
$N_s=576$ spins.
} \label{fig:stag} \ec
\end{figure}

We finish this investigation by calculating the N\'eel
temperature, asssuming a small (effective) 3D magnetic coupling
$\lambda_{3D}$ between the 2D planes. Using an RPA criterion, the
critical temperature $T_N$ is simply given by $\chi_{\rm
stag}(T_N)=1/|\lambda_{3D}|$ where the staggered spin
susceptibility (normalized per site) is defined as usual by,
\begin{equation}
\label{eq:RPA} \chi_{_{\rm stag}}(T)=\frac{1}{L^2} \sum_{i,j}
(-1)^{r_i + r_j} \int_0^\beta d\, \tau
\langle S_{i}^{z}(0)S_{j}^{z}(\tau)\rangle \, ,\nonumber
\end{equation}
and averaged over several disorder configurations (typically
$2000$). Since $\chi_{\rm stag}(T_N)$ is expected to reach its
thermodynamic limit for a {\it finite} linear size $L$ as long as
$T_N$ remains {\it finite}, accurate values of $T_N$ can be
obtained using a finite size computation of $\chi_{\rm stag}(T)$
for not too small inter-chain couplings. Fig.~\ref{fig:stag}(a)
shows that $\chi_{\rm stag}(T)$ diverges when $T\to 0$. $T_N$ is
determined by the intersection of the curve $\chi_{\rm stag}(T)$
with an horizontal line at coordinate $1/\lambda_{3D}$. Note that
finite size corrections remain small, even in the worst case
corresponding to very small $\lambda_{3D}$ values and large dopant
concentrations. The doping dependance of $T_N$ is plotted in
Fig.~\ref{fig:stag}(b) for a particularly small value
$\lambda_{3D}=0.01$ (in order to show the small size dependance
observable in that case). It clearly reveals a rapid decrease of
$T_N$ when $x\rightarrow 0$, but, in agreement with experiments,
does not suggest a non-zero critical concentration. In Fig.\ref{fig:stag}(b), 
we show the behavior of $T_N(x)$ down to $x\simeq0.007$. Note that from 
numerical fits of our data, we can not 
clearly distinguish between a power-law behavior (with an exponent $\sim 2.5$) 
and an exponential law like $A\exp(-B/x)$, as suggested by fits of experimental 
data for Cu$_{1-x}$Zn$_x$GeO$_3$ \cite{Manabe98}.
%$T_N\simeq 8.75\times x^{2.656}$ for $N_s=576$.

%%%%%%%%%%%FIG:TN%%%%%%%%%%%%%%%%%
%%%%%%%%%%%%%%%%%%%%%%%%%%%%%%%%%%%%%%%%%%%%

In summary, we have studied an effective low-energy Hamiltonian 
(valid for low temperature physics) describing the interaction of 
a finite concentration of spinless dopants randomly distributed in 
a generic low-dimensional (frustrated) SP system. 
The SSE method, which is applicable here because of the 
non-frustrated nature of the effective model, was used in combination
with finite-size scaling to compute both GS 
and finite temperature properties. The uniform susceptibility 
exhibits a Curie-like behavior down to very low temperature. 
We also predict that the AF order developps continuously without any finite 
critical dopant concentration in agreement with experiment.

AWS acknowledges support from the Academy of Finland, project
NO.~26175. D.P. thanks M.~Sigrist for pointing out that the 
uniform susceptibility data could be interpreted 
as a scaling regime. Further investigations along these lines are 
under way~\cite{scaling}. We thank IDRIS (Orsay, France) for using their supercomputer facilities.

\end{document}